\documentclass[conference]{IEEEtran}
\IEEEoverridecommandlockouts
\usepackage{cite}
\usepackage{amsmath,amssymb,amsfonts}
\usepackage{algorithmic}
\usepackage{graphicx}
\usepackage{textcomp}
\def\BibTeX{{\rm B\kern-.05em{\sc i\kern-.025em b}\kern-.08em
    T\kern-.1667em\lower.7ex\hbox{E}\kern-.125emX}}
\usepackage[utf8]{inputenc}
\usepackage{url}
\usepackage{svg}
\usepackage{tabularx}
\usepackage{textcomp}
\usepackage{lipsum}
\usepackage{xfrac}
\usepackage{commath}
\usepackage{makecell}
\usepackage{mathtools, nccmath}
\usepackage{multicol}
\usepackage[labelfont=bf]{caption} 
\usepackage{caption}
\usepackage[section]{placeins}
\usepackage{cleveref}
\usepackage {verbatim}
\usepackage{xcolor,colortbl}
\usepackage{ragged2e}
\usepackage{float}
\usepackage{blindtext}
\usepackage{cite}
\IEEEpubid{\makebox[\columnwidth]{978-1-6654-3156-9/21/\$31.00~\copyright{}2021 IEEE \hfill} \hspace{\columnsep}\makebox[\columnwidth]{ }}

\makeatletter
\newcommand{\specialcite}[1]{\@nameuse{b@#1}}
\makeatother
\begin{document}

\title{Bottlenecks in Blockchain Consensus Protocols}

\author{\IEEEauthorblockN{Salem Alqahtani}
\IEEEauthorblockA{\textit{Computer Science Department} \\
\textit{University at Buffalo,SUNY}\\
salemmoh@buffalo.edu}
\and
\IEEEauthorblockN{Murat Demirbas}
\IEEEauthorblockA{\textit{Computer Science Department} \\
\textit{University at Buffalo,SUNY}\\
demirbas@buffalo.edu}
}
\maketitle

\begin{abstract}

Most of the Blockchain permissioned systems employ Byzantine fault-tolerance (BFT) consensus protocols to ensure that honest validators agree on the order for appending entries to their ledgers. In this paper, we study the performance and the scalability of prominent consensus protocols, namely PBFT, Tendermint, HotStuff, and Streamlet, both analytically via load formulas and practically via implementation and evaluation. Under identical conditions, we identify the bottlenecks of these consensus protocols and show that these protocols do not scale well as the number of validators increases. Our investigation points to the communication complexity as the culprit. Even when there is enough network bandwidth, the CPU cost of serialization and deserialization of the messages limits the throughput and increases the latency of the protocols. To alleviate the bottlenecks, the most useful techniques include reducing the communication complexity, rotating the hotspot of communications, and pipelining across consensus instances.

\textbf{Keywords}: Consensus, PBFT, Tendermint, HotStuff, Streamlet, Byzantine fault-tolerance (BFT).
\end{abstract}
\section{Introduction}
\label{sec:introduction}

\IEEEPARstart{B}{lockchain} systems aim to provide trustless decentralized processing and storage of transactions, immutability, and tamper-resistance. Most of the Blockchains employ BFT~\cite{BFTGENERALS} consensus protocols to ensure that the validators agree on the order for appending new transactions to their ledgers. In particular, the Practical Byzantine Fault Tolerance (PBFT)~\cite{pbft} protocol forms the basis for most BFT consensus protocols, such as Tendermint~\cite{Tendermint}, Streamlet~\cite{streamlet}, and HotStuff~\cite{HOTSTUFF}.

PBFT builds on the Paxos~\cite{PAXOS} protocol and extends its crash failure to Byzantine fault-tolerance to defend against adversarial participants that can arbitrarily deviate from the protocol. PBFT upholds the safety of consensus with up to $1/3$ of the validators being Byzantine even in the asynchronous model, and maintains progress in a partially synchronous model. Since PBFT provides low latency, energy efficiency~\cite{quest}, and instant deterministic finality of transactions, PBFT is deemed suitable for many E-commerce applications that cannot tolerate long delays for transaction to be finalized and added to the ledger.

Unfortunately, the PBFT protocol has performance and availability problems. PBFT incurs quadratic message complexity and this curbs the scalability and performance of the consensus protocol. Secondly, PBFT leverages on a stable leader and changes it only if the leader is suspected to be Byzantine. Triggering a leader change requires a slow, costly, and prone to faults protocol which is called view change protocol.

To address these shortcomings of PBFT, blockchain systems mostly adopt rotating leader variants of PBFT. Tendermint~\cite{Tendermint} incorporates the leader rotation as part of the normal consensus path. While this adds some cost in terms of performance, it pays off in terms of fault-tolerance, availability, and fairness.

Streamlet~\cite{streamlet} gives a two-phase rotating leader solution avoiding a lot of overhead in Tendermint. HotStuff~\cite{HOTSTUFF} incorporates pipelining to rotation of leaders to improve throughput further. It also addresses the quadratic message complexity in PBFT and Tendermint, and provides a responsive protocol with linear complexity.

Although these rotating leader variants improve on PBFT, there has not been any study to investigate how they compare with each other and how effective different strategies for leader rotation are for alleviating bottlenecks in BFT protocols.

\begin{figure*}[htb]
\centering
\vspace*{-4mm}
\includegraphics[width=\textwidth,height=3cm]{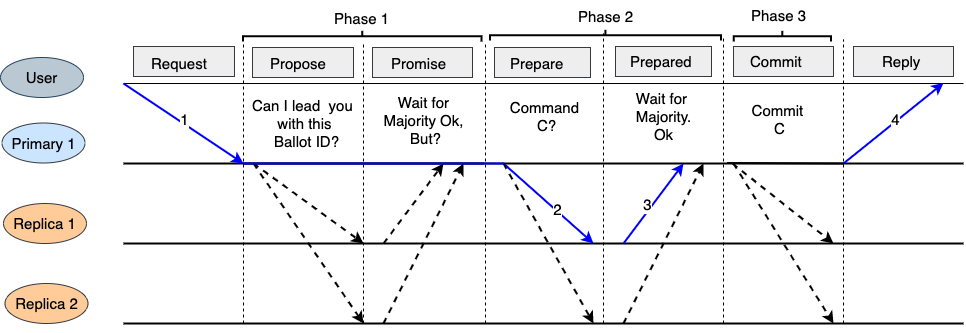}
\vspace*{-7mm}
\caption{\textbf{Paxos protocol}}
\label{fig:PAXOS}
\end{figure*}

\par
\textbf{Contributions.}
In this paper, we provide a comprehensive systematic investigation of bottlenecks in deterministic finality BFT consensus protocols, namely PBFT, Tendermint, HotStuff, and Streamlet. 

We take a two-pronged approach. First, we provide a theoretical analysis of complexity of these consensus protocols. Second, we build a novel framework called PaxiBFT~\cite{PaxiBFT}. The purpose of the PaxiBFT framework is to implement, benchmark, and evaluate BFT protocols performance under identical conditions. PaxiBFT is written in Go and built in modules that makes it easy for developers to modify and evaluate their own protocols. On PaxiBFT~\cite{PaxiBFT}, we built and evaluated Paxos~\cite{pbft}, PBFT\cite{pbft}, Tendermint~\cite{Tendermint}, Streamlet~\cite{streamlet}, and HotStuff~\cite{HOTSTUFF}.

We study the bottlenecks of these consensus protocols and identify the factors that limit their scalability. Our investigations point to the communication complexity as the culprit. Even when there is enough network bandwidth, the CPU cost of serialization and deserialization of the messages limits the throughput and increases the latency of the protocols. We find that HotStuff performs significantly better than the other protocols because it
(1) replaces all-to-all communication with all-to-one communication,
(2) rotates the leaders at the hotspot of all-to-one communication across rounds to shed and balance load, and
(3) employs pipelining across rounds to improve throughput further.

Our analysis and evaluation about the bottlenecks can pave the way for designing more efficient protocols that alleviate the identified performance bottlenecks. These analysis and evaluation results will also help researchers and developers to choose suitable consensus protocols for their needs. 

\vspace{3mm}

{\bf Outline of the rest of the paper.} After discussing the background and related work, we explain distributed consensus in Section~\ref{sec:back}, and present rotated leader BFT consensus protocols in Second~\ref{sec:back1}. We analyze the protocols in Section~\ref{sec:anal}. We discuss our implementations in Section~\ref{sec:rw} and present evaluation results in Section~\ref{sec:eval}. 

\section{Background and Related Work}
\label{sec:permodel}

\subsection{Related Work}

A plethora of surveys on BFT consensus protocols in the permissioned model have come out recently, which focus on their comparisons on theoretical results. The survey~\cite{15} states that there is no perfect consensus protocol and presents their trade-offs among security and performance. A recent survey~\cite{1} provides an overview of the consensus protocols used in permissioned blockchain and investigates the algorithms with respect to their fault and resilience models. Another work~\cite{16} investigates the relationship between blockchain protocols and BFT protocols. A more recent work~\cite{7} classifies consensus protocols as proof-based and vote-based, and argues that vote-based protocols are more suitable for permissioned blockchain whereas proof of work/stake/luck based protocols are more suitable for public blockchains. There have been more exhaustive theoretical surveys such as~\cite{bano2019sok} on committee and sharding based consensus protocols. The work summarized variants of protocols, their challenges, and both their designs and their security properties.

While there has been a lot of work on consensus protocols, there has not been any work for evaluating and analyzing the performance bottlenecks in these consensus protocols. This is due to the fact that consensus protocols are more complex and not easy to implement. Motivated by this fact, we evaluate the performance of consensus protocols with finality property that work in a partial synchrony model.

\section{Canonical Consensus Protocols}
\label{sec:back}
Paxos is widely used in research and in practice to solve decentralized consensus. Unlike the crash failure model in Paxos, the byzantine failure model is more complex and uses a number of cryptographic operations. As our best case scenario to compare consensus protocols performances, we have chosen Paxos as a performance bar to compare with other protocols instead of Raft~\cite{ongaro2014search} which uses in Hyperledger Fabric and has the same performance as Paxos~\cite{dtm}. 
\subsection{Paxos}
Paxos protocol~\cite{PAXOS} was introduced for achieving consensus among a set of validators in an asynchronous setup prone to crash failures. Paxos requires at least $N\!\geq\!2F\!+\!1$ validators to tolerate the failure of $F$ validators. By using majority quorums, Paxos ensures that there is at least one validator in common from one majority to another, and avoids the split-brain problem. 
\begin{figure*}[htb]
	\centering
	\vspace*{-4mm}
	  \includegraphics[width=\textwidth,height=3.5cm]{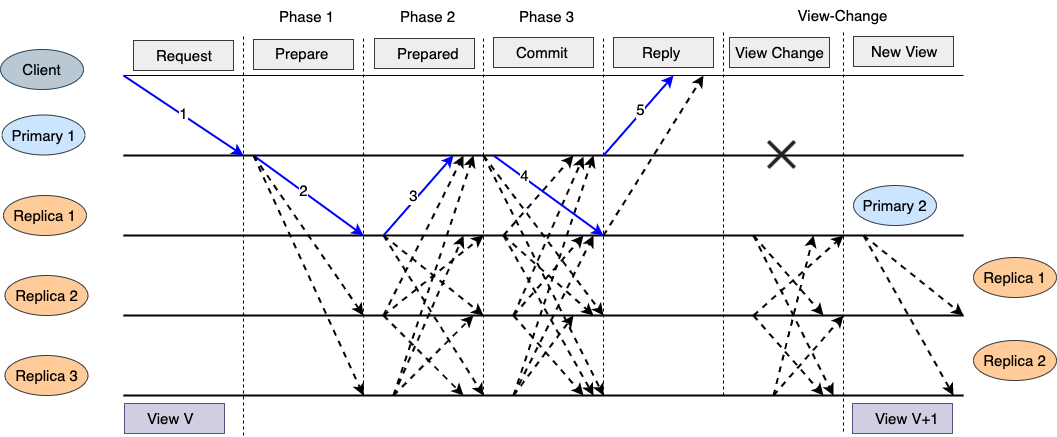}
	  \vspace*{-8mm}
	\caption{\textbf{Practical byzantine fault tolerance protocol}}
	\label{fig:PBFT}
\end{figure*} 
\subsubsection*{The Protocol}
Paxos architecture is illustrated in Figure~\ref{fig:PAXOS}.
\begin{itemize}
\item[$\ast$] A candidate leader tries to become the leader by starting a new round via broadcasting a propose message with its unique ballot number $bal$. The other validators acknowledge this propose message with the highest ballot they have seen so far, or reject it if they have already seen a ballot number greater than $bal$. Receiving any rejection fails the candidate leader. 
\item[$\ast$] After collecting a majority quorum of acknowledgments, the candidate leader becomes the leader and advances to the prepare phase, where the leader chooses a value for its ballot. The value would be the value associated with the highest ballot learned in the previous phase. In the absence of any such pending proposal value,  a new value is chosen by the leader. The leader asks its followers to accept the value and waits for the acknowledgment messages. Once the majority of followers acknowledge the value, it becomes anchored and cannot be revoked. Again a single rejection message nullifies the prepare phase, revokes leadership of the node, and sends it back to propose phase it cares to contend for the leadership. 
\item[$\ast$] Upon successful completion of the prepare phase, the leader node broadcasts a commit message in the commit phase. This informs the followers that a majority quorum accepted the value and anchored it, so that the followers can also proceed to commit the value.
\end{itemize}

\begin{figure*}[!htb]
	\centering
	\vspace*{-4mm}
	  \includegraphics[width=\textwidth,height=3.5cm]{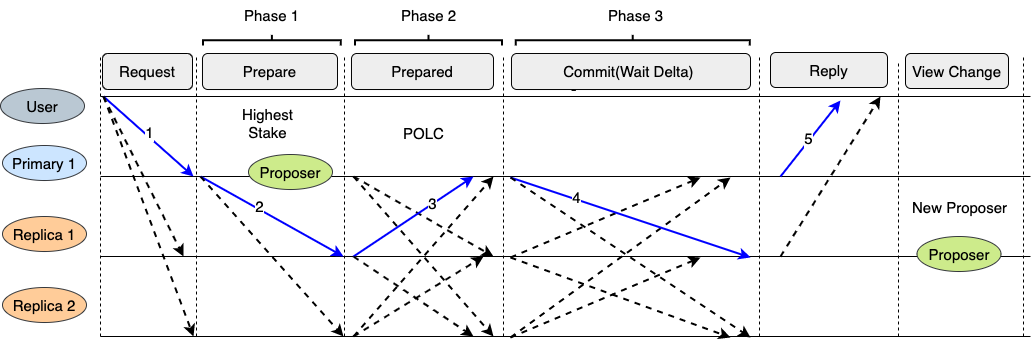}
	  \vspace*{-7mm}
	\caption{\textbf{Tendermint protocol}}
	\label{fig:TM}
      \end{figure*}
      
\subsection{PBFT}
\label{sec:pbft}
PBFT protocol~\cite{pbft} provided the first practical solution to the Byzantine problem. PBFT employs an optimal bound of $\!N\!\!\!\geq\!\!3F\!\!+\!\!1\!$ validators, where the Byzantine adversaries can only control up to $\!F\!$ validators. PBFT uses encrypted messages to prevent spoofing and replay attacks, as well as detecting corrupted messages. PBFT employs a leader-based paradigm, guarantees safety in an asynchronous model, and guarantees liveness in a partially synchronous model. When the normal path does not make progress, PBFT uses a view change protocol to elect a new leader.

\subsubsection*{The Protocol}
PBFT architecture is illustrated in Figure~\ref{fig:PBFT}.

\begin{itemize}
\item[$\ast$] The leader receives the encrypted client's request and starts its prepare phase by proposing the client's request along with its view number to all followers. The followers broadcast the client's request either to acknowledge the leader or reject it if they have already seen a higher view number.

\item[$\ast$] In the absence of a rejection, each follower waits for $\!N\!-\!F\!$ matching prepared messages. This ensures that the majority of correct validators has agreed on the sequence and view numbers for the client's request.

\item[$\ast$] The followers advance to the commit phase, re-broadcast the proposal, and waits for $\!N\!-\!F\!$ matching commit messages. This guarantees the ordering across views.

\item[$\ast$] Finally, $\!F\!+\!1\!$ validators reply to the client after they commit the value.

\end{itemize}
In case of a faulty leader, a view-change protocol is triggered by the non-faulty validators that observe timer expiration or foul play. Other validators join the view change protocol if they have seen $F\!+\!1$ votes for the view change and the leader for the next view tries to take over. The new leader must decide on the latest checkpoint and ensure that non-faulty validators are caught up with the latest states. View change is an expensive and bug-prone process for even a moderate system size.

\section{Rotated Leader Protocols}
\label{sec:back1}
In this section, we provide an overview of Tendermint, Tendermint*, Streamlet, and HotStuff BFT protocols.

\subsection{Tendermint BFT}
Tendermint protocol~\cite{Tendermint}, used by Cosmos network~\cite{cosmos}, utilizes a proof-of-stake for leader election and voting on appending a new block to the chain. Tendermint rotates its leaders using a predefined leader selection function that priorities selecting a new leader based on its stake value. This function points to a proposer responsible for adding the block in blockchain. The protocol employs a locking mechanism after the first phase to prevent any malicious attempt to make validators commit different transactions at the same height of the chain. Each validator starts a new height by waiting for prepare and commit votes from $2F+1$ validators and relies on the gossip network to spread votes among all validators in both phases.

Tendermint prevents the hidden lock problem~\cite{Tendermint} by waiting for $\delta$ time. The hidden lock problem occurs because receiving $\!N\!-\!F\!$ replies from participants (up to $\!F\!$ of which may be Byzantine) alone is not sufficient to ensure that the leader gets to see the highest lock; the highest lock value may be hidden in the other $\!F\!$ honest nodes which the leader did not wait to hear from. Such an impatient leader may propose a lower lock value than what is accepted and this in turn may lead to a liveness violation. The rotation function that elects a next leader enables Tendermint to skip a faulty leader in an easy way that is integrated to the normal path of the protocol.

\begin{figure*}[!htb]
	\centering
	\vspace*{-4mm}
	  \includegraphics[width=\textwidth,height=3.5cm]{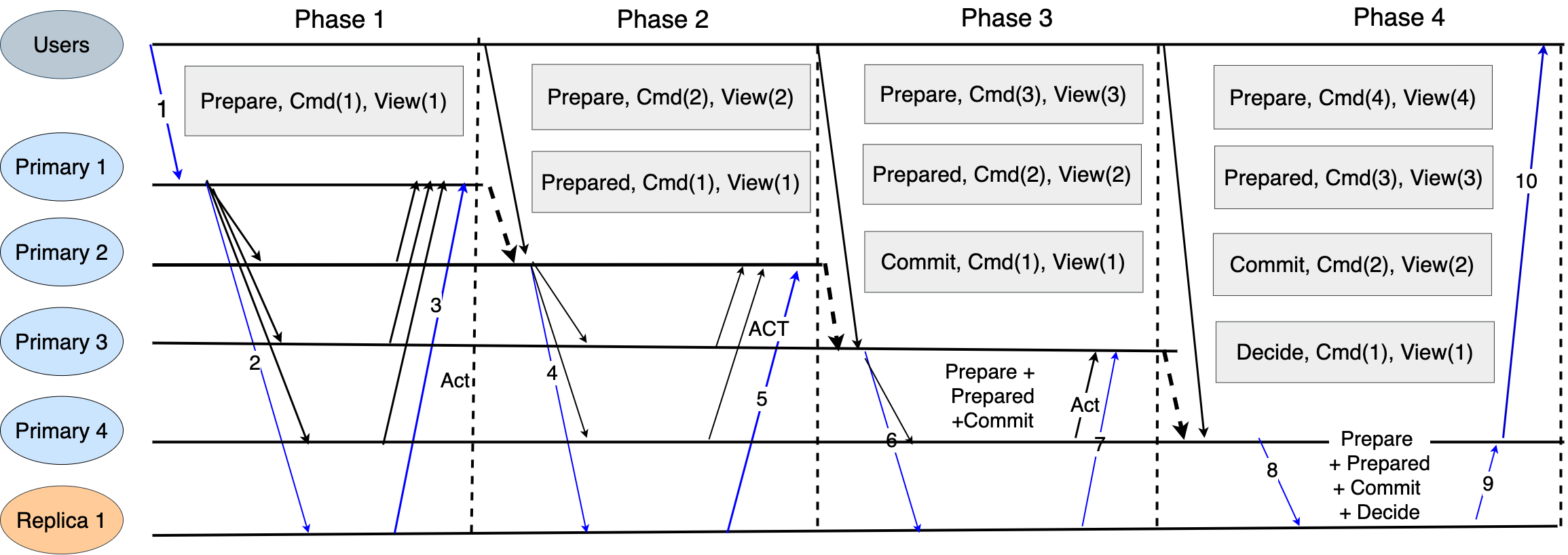}
	  \vspace*{-7mm}
	\caption{\textbf{HotStuff protocol}}
	\label{fig:HotStuff}
\end{figure*}

\subsubsection*{The Protocol}
Tendermint protocol is illustrated in Figure~\ref{fig:TM}.

\begin{itemize}
\item[$\ast$] A validator becomes a leader if it has the highest stake value. It starts the prepare phase by proposing the client's request to all followers. Followers wait $\delta$ time for the leader to propose the value of the phase. If the followers find that the request came from a lower height than their current blockchain height, or that they did not receive any proposal from the leader, they gossip a nil block. Otherwise, the followers acknowledge the leader's request, then gossip the request and prepared message to other nodes.

\item[$\ast$] Upon receiving a majority of prepared messages in the prepared phase, a node locks on the current request and gossips a commit message. Otherwise, a follower rejects the prepared value and gossips the previous locked value.

\item[$\ast$] Upon receiving the majority votes in the commit phase, the nodes commit the value and reply to the client's request. Otherwise, they vote nil.

\item[$\ast$] If the leader is able to finish the view and commit the block, all validators move to the next height of the chain.
\end{itemize}

\begin{figure*}[!htb]
	\centering
	\vspace*{-5mm}
	  \includegraphics[width=\textwidth,height=2.7cm]{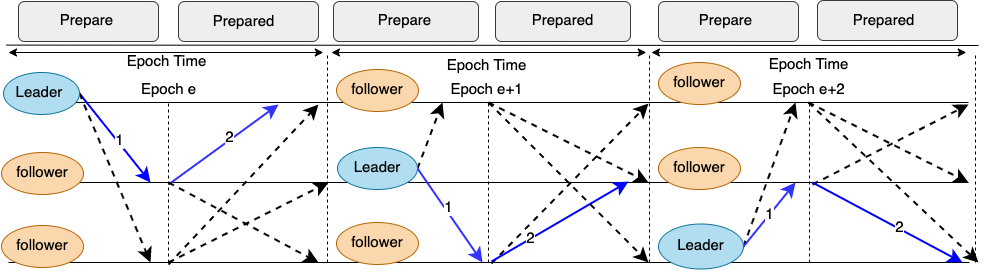}
	  \vspace*{-7mm}
	\caption{\textbf{Streamlet protocol}}
	\label{fig:streamlet}
\end{figure*}

Tendermint* is a hypothetical variant of Tendermint we consider for evaluation purposes. It differs from Tendermint only in two parts. It forgoes the $\delta$ time in commit phase and the all-to-all communication in Tendermint, replacing that instead with a direct communication with just the leader. Even though the protocol violates correctness properties of BFT, we employ it in order to demonstrate which components of the protocols are responsible for how much performance gains/penalties and explore these in Sections~\ref{sec:eval} and~\ref{sec:anal}.

\subsection{HotStuff BFT}
\label{sec:hs}
HotStuff protocol~\cite{HOTSTUFF}, is used in Facebook's Libra~\cite{Librag}. HotStuff rotates leaders for each block using a rotation function. HotStuff is responsive; it operates at network speed by moving to the next phase after the leader receives $N-F$ votes. This is achieved by adding a pre-commit phase to the lock-precursor. To assign data and show proof of message reception and progression, the protocol uses Quorum Certificate(QC), which is a collection of $N-F$ signatures over a leader proposal. Moreover, HotStuff uses one-to-all communication. This reduces the number of message types and communication cost to be linear. The good news is that, since all phases become the same communication-pattern, HotStuff uses pipeline mechanism and performs four leader blocks in parallel; thus improving the throughput by four. 

\subsubsection*{The Protocol}
HotStuff protocol is illustrated in Figure~\ref{fig:HotStuff}. 
\begin{itemize}

\item[$\ast$] A new leader collects new-view messages from $\!N\!-\!F\!$ followers and the highest prepare QC that each validator receives. The leader processes these messages and selects the prepare QC with the highest view. Then, the leader broadcasts the proposal in a prepare message.

\item[$\ast$] Upon receiving the prepare message from the leader, followers determine whether the proposal extends the highest prepare QC branch and has a higher view than the current one that they are locked on. 

\item[$\ast$] The followers send acknowledgement back to the leader, who then starts to collect acknowledgements from $\!N\!-\!F\!$ prepare votes. Upon receiving $\!N\!-\!F\!$ votes, the leader combines them into a prepare QC and broadcasts prepare QC in pre-commit messages.

\item[$\ast$] A follower responds to the leader with a pre-commit vote. Upon successfully receiving $\!N\!-\!F\!$ pre-commit votes from followers, the leader combines them into a pre-commit QC and broadcasts them in commit messages. 

\item[$\ast$] Followers respond to the leader with commit votes. Then, followers lock on the pre-commit QC. Upon successfully receiving $N-F$ commit votes from followers, the leader combines them into a commit QC and broadcasts the decide messages.

\item[$\ast$] Upon receiving a decide message, the followers execute the commands and start the next view.

\end{itemize}

HotStuff pipelines the four phase leader-based commit to a pipeline depth of four, and improves the system throughput to commit one client's request per phase. As per this pipelining, each elected leader proposes a new client request on every phase in a new view for all followers. Then, the leader simultaneously piggybacks pre-commit, commit, and decide messages for previous client requests passed on to it from the previous leader through commit certificate.

\subsection{Streamlet BFT}

Streamlet protocol proposed in 2020~\cite{streamlet}. Streamlet leverages the blockchain infrastructure in addition to the longest chain rule in Nakamoto protocol~\cite{bitcoin} to simplify consensus. Streamlet rotates its leader for each block using a rotation function. The protocol proceeds in consecutive and synchronized epochs where each epoch has a dedicated leader known by all validators. Each epoch has a leader-to-participants and participants-to-all communication pattern. This reduces the number of message types, but the communication cost is $O(N^3)$. Streamlet has a single mode of execution and there is no separation between the normal and the recovery mode. Streamlet guarantees safety even under an asynchronous environment with arbitrary network delays and provides liveness under synchronous assumptions.

\subsubsection*{The Protocol}
Streamlet protocol is illustrated in Figure~\ref{fig:streamlet}. 

\begin{itemize}

\item[$\ast$] The candidate leader for epoch($e_{i}$) broadcasts a block that extends the longest finalized blockchain it has seen. 

\item[$\ast$] Upon receiving propose message from the leader, validator nodes acknowledge the proposed block with the highest view number and the longest chain that they have seen so far. Then validator nodes broadcasts a vote message in the vote phase.

\item[$\ast$] Both leader and followers collect a majority quorum of acknowledgments equals to $\!2N\!/3\!$ for the proposal block in epoch($e_{i}$) and mark the block as notarized block.

\item[$\ast$] If a validator node finds three consecutive notarized blocks in the blockchain($e_{i},e_{i+1},e_{i+2}$), the validator node finalize up the chain. 
\end{itemize}

\section{Analysis and Discussion}
\label{sec:anal}
In this section, we compare the strengths and weaknesses of the consensus protocols considered and provide back-of-the-envelope calculations for estimating performance.

\subsection{Theoretical analysis}
Table~\ref{tab:template} provides a synopsis of the blockchain protocols characteristics we studied. We elaborate on these next.

\begin{table*}
\vspace*{-4mm}
\centering
\begin{tabular}{ |l|c|c|c|c|c|c| } 
 \hline
&Paxos~\cite{PAXOS}&PBFT~\cite{pbft}&Tendermint~\cite{Tendermint}&Tendermint*~\cite{Tendermint}&HotStuff~\cite{HOTSTUFF}&Streamlet~\cite{streamlet}\\
\hline
Synchrony &  \multicolumn{6}{|c|}{Partially synchronous}\\
\hline
Communicating Node & Centralized & Broadcast & Gossip & Centralized & Centralized & Broadcast \\
\hline
Critical Path Messages & 4 & 5 & 5 & 8 & 10& 4\\
\hline
Normal Message Complexity & $O(N)$ & $O(N^2)$ & $O(N^2)$   & $O(N)$   & $O(N)$  &  $O(N^3)$\\
\hline

Multiple View Change & $O(N^2)$ & $O(N^4)$ & $O(N^3)$& $O(N^2)$ & $O(N^2)$&$O(N^4)$\\
\hline
Responsive & Yes & Yes & No & Yes & Yes& No \\
\hline
\end{tabular}
\caption{\textbf{Characteristics of BFT consensus protocols}}
\label{tab:template}
\vspace*{-3mm}
\end{table*}

\textbf{Synchrony Requirements.}
All protocols that we considered assume partially synchronous network model~\cite{ps}. In this model, after a period of asynchrony, the network starts to satisfy synchrony assumptions and honest messages will be delivered within the synchronous period. 

\textbf{Time Complexity.} PBFT normal execution has a quadratic complexity. When the leader is a malicious, the protocol changes the view with a different leader using a view-change which contains at least $2F+1$ signed messages. Then, a new leader broadcasts a new-view message including the proof of $2F+1$ signed view-change messages. Validators will check the new-view message and broadcast it to have a match of $2F+1$ new-view message. The view-change has then $O(N^3)$ complexity and $O(N^4)$ in a cascading failure~\cite{dtm}.

Tendermint reduces PBFT's message complexity to $O(N^3)$ in the worst case. Since at each epoch all validators broadcast messages, the protocol uses $O(N^2)$ messages. Thus, in the worst case scenario when there is $F$ faulty validators, the message complexity is $O(N^3)$~\cite{dtm}.

Paxos, Tendermint*, and HotStuff all have linear message complexity. The worse case cost in these protocols is $O(N^2)$ considering worst-case consecutive view-changes.

Streamlet has message complexity $O(N^3)$. Streamlet loses linear communication complexity due to all-to-all communication in vote message. In the worst case when there is a leader cascading failure, the Streamlet complexity is $O(N^4)$.

All of the protocols provide responsiveness except for the Tendermint due to $\delta$ waiting time in commit phase and for the Streamlet due to its fixed epoch length.

\subsection{Load and Capacity}

Our considered protocols reach consensus once a quorum of participants agrees on the same decision. A quorum can be defined as sets containing majority validators in the system with every pairs of set has a non-empty intersection. To select quorums $Q$, quorum system has a strategy $S$ in place to do that. The strategy decides which quorums types to choose that leads to a load on each validator. The load $\ell(S)$ is the minimum load on the busiest validator. The capacity $Cap(S)$ is the highest number of quorum accesses that the system can possibly handle $Cap(S)=\frac{1}{\ell(S)}$~\cite{quorum}. 

In single leader protocols, the busiest node is the leader~\cite{dissectingPaxos}. \useshortskip

\begin{equation}\label{eq1}
\begin{split}
\ell(S) & = \frac{1}{L}(Q-1)NumQ+(1-\frac{1}{L})(Q-1)NumQ\\
 \end{split}
\end{equation}

where $Q$ is the quorum size chosen in both leader and followers, NumQ is quorums number handled by leader/follower for every transaction, and $L$ is the number of operation leaders. There is a $\frac{1}{L}$ chance the validator is the leader of a request. Leader communicates with $N-1=Q$ validators and we assume $N = 9$. The probability of the node being a follower is $1- \frac{1}{L}$, where it only handles one received message in the best case. The protocols perform better as the load decreases.\useshortskip
\begin{equation}\label{eq2}
\begin{split}
\ell(Paxos) & = 4
\end{split}
\end{equation}
\useshortskip
In Paxos, equation~\ref{eq2} with L = 1, quorum size Q = $\lfloor{\frac{N}{2}}\rfloor$ + 1, and number of quorums $NumQ = 1$.
\useshortskip
The equation~\ref{eq3} is a PBFT protocol with Q = $\lfloor{\frac{2*N}{3}}\rfloor$, and $NumQ=2$.\useshortskip

\begin{equation}\label{eq3}
\begin{split}
\ell(PBFT) & = 10
\end{split}
\end{equation}

The equation~\ref{eq3}, PBFT~\ref{sec:pbft} has high load which implies that the throughput is low. In Section~\ref{sec:eval}, our evaluation illustrates how low throughput is comparing to other protocols. This is an indication how load is related to the throughput in our equation~\ref{eq1}. PBFT bottleneck becomes quicker fast due to high load that comes form all-to-all communications.\useshortskip

The equation~\ref{eq4} is a rotated leader HotStuff protocol with a leader Q = $\lfloor{\frac{2*N}{3}}\rfloor$, $NumQ=4$, and $L = 4$. Unlike PBFT, HotStuff followers have no quorums. So, the $NumQ=0$ in the followers nodes.

\begin{equation}\label{eq4}
\begin{split}
\ell(HotStuff)& =  5
\end{split}
\end{equation}

The equation~\ref{eq4}, HotStuff~\ref{sec:hs} has lowest load which implies that the throughput is high. In Section~\ref{sec:eval}, our evaluation illustrates how high throughput is comparing to other protocols. This is an indication how load is related to the throughput in our equation~\ref{eq1}. HotStuff bottleneck did not grow fast due to low load that comes form one-to-all communications and pipeline techniques.

Tendermint has $\delta$ waiting time before committing the value and Streamlet is a synchronous clock. We eliminate them from our load analysis because busiest node affected not by actual workload but also by waiting time. 

\subsection{Latency}
The formula~\ref{eq5} calculates the latency of consensus in the protocols considered, except for Streamlet which has a fixed epoch time due to its synchronous clock for each instance of consensus.\useshortskip
\begin{equation}
\begin{split}\label{eq5}
Latency(S) = Critical\; Path  + D_{L} + \delta
\end{split}
\end{equation}

Critical Path is the round trip message between a designated leader and its followers. Paxos's critical path has a 2-message delay as illustrated in Figure~\ref{fig:PAXOS}. With the help of a stable leader, Paxos reduces message latency in the first phase. $D_{L}$ is the round trip message between a client and designated leader. In Table~\ref{tab:template}, PBFT and Tendermint have a 5-message delay as illustrated in Figures~\ref{fig:PBFT} and~\ref{fig:TM}. Paxos and Streamlet have a 4-message delay. $\delta$ refers to the waiting time that the leader has to wait before committing transactions.

As the number of validators increases, bottlenecks arise and the above latency formula starts to break down, as we see in Section~\ref{sec:eval}. The reasons are different communication patterns along with different loads imposed on protocols.

\section{PaxiBFT Framework}
\label{sec:rw}
\begin{figure}[t!]
\centering
\vspace*{-4mm}
\includegraphics[width=\linewidth]{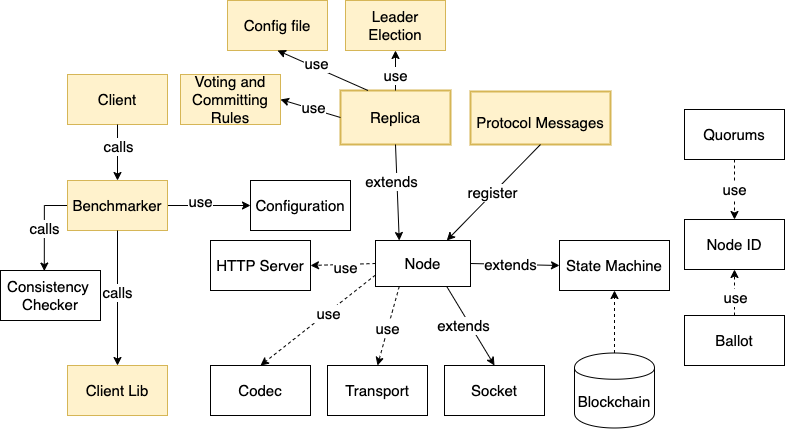}
\vspace*{-6mm}
\caption{\textbf{The PaxiBFT architecture}}
\label{fig:paxi}
\end{figure}

Our experiments are performed on the PaxiBFT~\cite{PaxiBFT} framework \footnote{https://github.com/salemmohammed/PaxiBFT} written in Go. PaxiBFT enables evaluation of BFT consensus protocols and supports both customization of workloads and deployment conditions. The PaxiBFT architecture is shown in Figure~\ref{fig:paxi}. The PaxiBFT's purpose is to offer a fair environment for comparing BFT protocols. 

To implement BFT consensus protocols in PaxiBFT framework, we designed BFT client library, benchmarker, message handling modules. For the network infrastructures, we borrowed the core network implementations from Paxi framework~\cite{PAXI}. The client library can send a request to all validators and receive $\!F\!\!+\!\!1\!$ replies. We also enable the benchmark to be able to measure the latency for each request. 

In PaxiBFT, all BFT protocols can be implemented by coding the protocols' phases, functions, and message types. In Figure~\ref{fig:paxi}, we highlighted some important components that can be modified by developers to implement new BFT protocol.

The top layer of PaxiBFT framework consists of config file, message file, and validator code. The config file is distributed among all validators in JSON format, which contains all validator addresses, quorum configurations, buffer sizes, networking parameters, and benchmark parameters. The developers specify the message structures that need to be sent between validators in the message file. Finally, in the validator file, the developers write the code to handle client requests and implement the replication protocol.

In lower layer of PaxiBFT framework, the core network network implementations as we mentioned earlier borrowed from Paxi framework~\cite{PAXI}. The networking interface encapsulates a message passing model, exposes basic APIs for a variety of message exchange patterns, and transparently supports TCP, UDP, and simulated connection with Go channels. The Quorums interface provides multiple types of quorum systems. The key-value store provides an in-memory multi-version key-value datastore that is private to every node. The client library uses a RESTful API to interact with any system node for read and write requests. This allows users to run any benchmark (e.g. YCSB~\cite{ycsb}) against their implementation in Paxi without porting the client library to other programming languages. Finally, the benchmarker component generates workloads with tunable parameters for evaluating performance and scalability.
\section{Experimental Results}
\label{sec:eval}
\subsection{Experimental Setup}
The experiments were conducted on AWS instances EC2 m5a.large, with 2 vCPU. The experiments were performed with network sizes of 4 to 20 nodes. Based on our experiments results in Section~\ref{sec:evalres}, this network size is appropriate to state and conclude our findings. To push system throughput, we varied the number of clients up to 90 and used a small message size. In our experiments, message size did not dominate consensus protocols performance, but the complexity of consensus protocols dominates the performance. We defined the throughput as the number of transactions per second (tx/s for short) that validator processes. We conducted our experiments in LAN deployment and Wide Area Network(WAN) across 4 AWS regions(Ohio, N.California, Oregon, and N.Virginia). In WAN, pushing the system throughput to its limit to get the system bottlenecks was difficult while it was easy in LAN due to the short network pipe between instances.

In Tendermint, as we discussed in Section~\ref{sec:back1}, waits $\!\delta\!$ time before committing the block to solve hidden lock problem. This $\!\delta\!$ time includes one way message time and committing time. In Streamlet protocol, as we discussed in Section~\ref{sec:back1}, the epoch time includes round trip communication time and propose-vote computing time. In LAN, We set $\!\delta\!$ time in Tendermint to be 2 ms and epoch time in Streamlet to be 3 ms. In WAN, We set $\!\delta\!$ and epoch time in Streamlet to be 50 ms. Our experiments show that these choices of $\!\delta\!$ and epoch durations are sufficient and ensure safe execution of both protocols.

\subsection{Evaluation Results}
\label{sec:evalres}
\begin{figure}[!h]
	\centering
	\vspace*{-5mm}
	\includegraphics[width=3.8in]{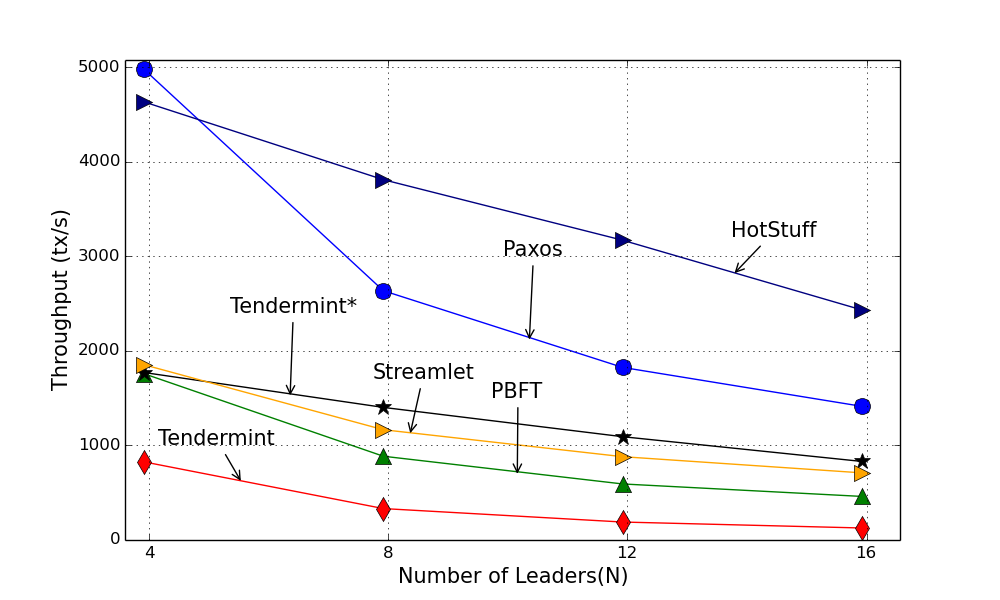}
	\vspace*{-8mm}
	\caption{\textbf{Throughput comparison in LAN}}
	\label{fig:20TC}
\end{figure}

\begin{figure}[!h]
	\centering
	\vspace*{-8mm}
	\includegraphics[width=3.8in]{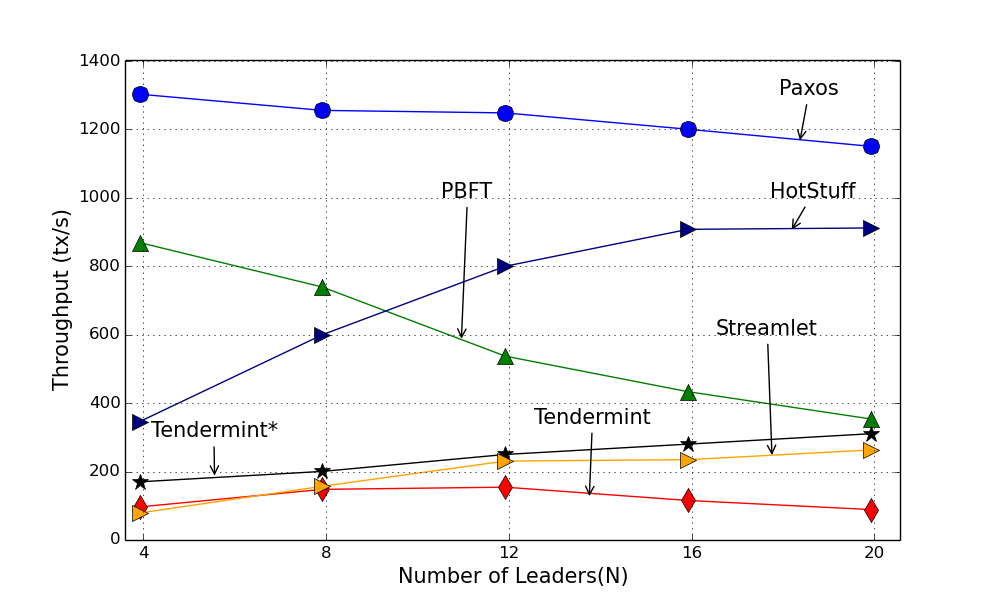}
	\vspace*{-8mm}
	\caption{\textbf{WAN’s throughput comparison in Virginia, California, Oregon, and Ohio}}
	\label{fig:20TCw}
\end{figure}

{\bf Paxos.} We evaluated Paxos as our baseline system. Figure~\ref{fig:20TC} and Figure~\ref{fig:20TCw} show that Paxos throughput declines as we increase the number of validators $\!N\!$. For example, when $\!N\!$ is 4 and clients are 90, the number of transactions that the system can process is approximately 4900 tx/s. On the other hand, when $\!N\!$ equals to 16, with the same number of clients, the system can only handle 1500 tx/s. This is due to the communication bottleneck at the single leader in Paxos~\cite{dissectingPaxos}. The Paxos experimental result demonstrates that the load on single leader increased significantly which matches our loading Formula~\ref{eq2}.

Latency increases as $\!\!N\!\!$ is increased because the leader struggles to communicate with more validators due to the cost of CPU being utilized in serialization/deserialization of messages.

{\bf PBFT.} The throughput evaluation is shown in Figure~\ref{fig:20TC} and Figure~\ref{fig:20TCw}. The all-to-all communication leads to a substantial throughput penalty. PBFT is also limited by a single leader communicating with the clients. When $\!N\!$ is 4 and clients are 90, the number of transactions that the system can process is around 1750 tx/s in LAN and 870 tx/s in WAN. However, with the same number of clients, and $\!N\!=\!16$, the system can only handle around 500 tx/s and 350 tx/s in WAN. The PBFT experimental result shows how significant the performance bottlenecks become in comparison to Paxos. Theoretically, we captured this high load in PBFT loading Formula~\ref{eq3}.

{\bf Tendermint.} Throughput results are shown in Figure~\ref{fig:20TC} and Figure~\ref{fig:20TCw}. The clients are configured to communicate with all validators for all operations. Tendermint performance is bad because the protocol inherits all of the PBFT bottlenecks and tops them with waiting maximum network delay $\!\delta\!$ for solving hidden lock problem. For $\!N\!=\!16$, Tendermint degrades to 150 tx/s in LAN and around 90 tx/s in WAN.

\begin{figure}[!h]
	\centering
	\vspace*{-8mm}
	\includegraphics[width=3.8in]{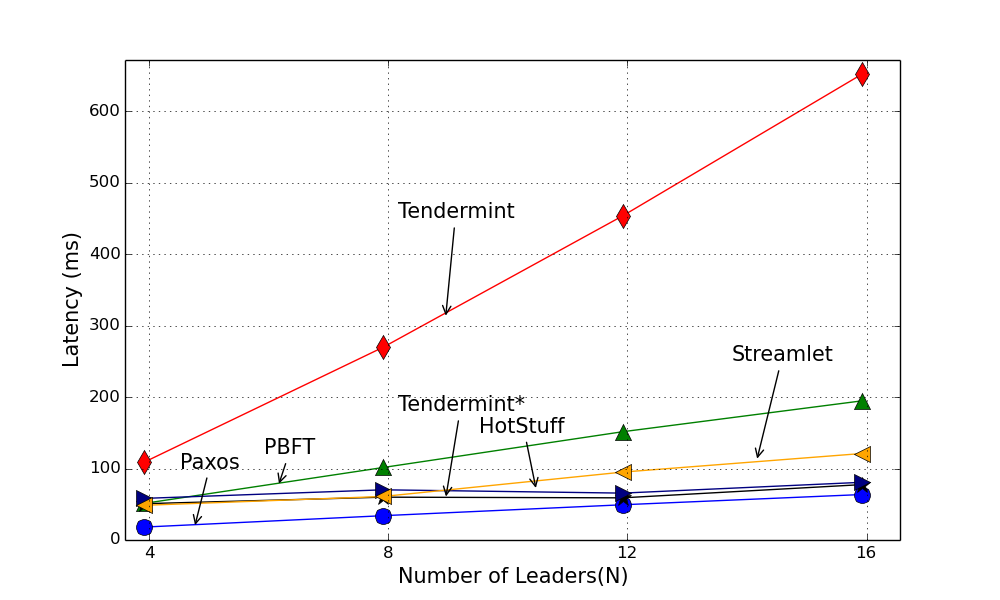}
	\vspace*{-8mm}
	\caption{\textbf{Latency comparison}}
	\label{fig:20LC}
\end{figure}

{\bf Tendermint*.} The throughput is shown in both Figure~\ref{fig:20TC} and Figure~\ref{fig:20TCw}, and latency in Figure~\ref{fig:20LC}. Tendermint* is a hypothetical protocol that waives the all-to-all communication and the $\!\delta\!$ time delay in Tendermint for evaluation/comparison purposes to identify those overheads. As such we can see that there is around 4 times improvement in throughput and latency in Tendermint* as compared to Tendermint.

{\bf HotStuff.} HotStuff achieves the best throughput compared to the other protocols, as shown in Figure~\ref{fig:20TC} and Figure~\ref{fig:20TCw}. This is because HotStuff uses leader-to-all and all-to-leader communication, as in Paxos, and introduces pipelining of 4 different leaders' consensus slots. Compared to PBFT and Tendermint, HotStuff enables pipelining due to normalizing all the phases to have the same structure. It also adds an additional phase to each view, which causes a small amount of latency, and allows HotStuff to avoid the $\!\delta\!$ waiting time.

{\bf Streamlet.} The maximum throughput is around 700 tx/s with $\!epoch\!=\!3\!$ ms while 300 tx/s with $\!epoch\!=\!50\!$ ms in WAN. The synchrony clock, all-to-all communication in the second phase, and the lack of pipeline techniques result in a substantial loss in the protocol's throughput. On the other hand, the Streamlet protocol has only one phase (propose and vote), which simplifies its architecture.

\begin{figure}[!h]
	\centering
	\vspace*{-8mm}
	\includegraphics[width=3.8in]{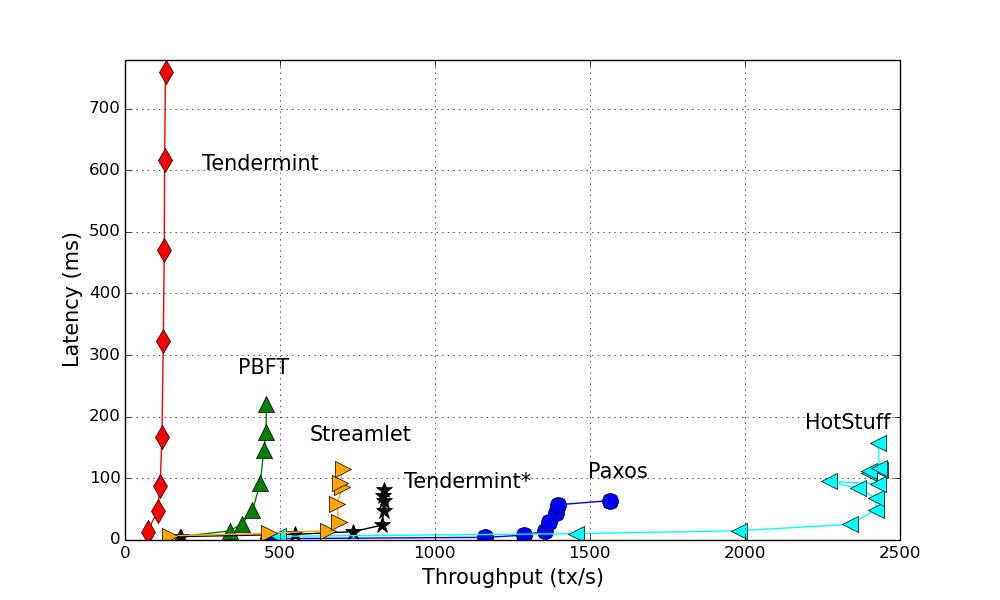}
	\vspace*{-8mm}
	\caption{\textbf{System throughput and the latency on 20-node LAN cluster}}
	\label{fig:LvT}
\end{figure}

\begin{figure}[!h]
	\centering
	\vspace*{-6mm}
	\includegraphics[width=3.8in]{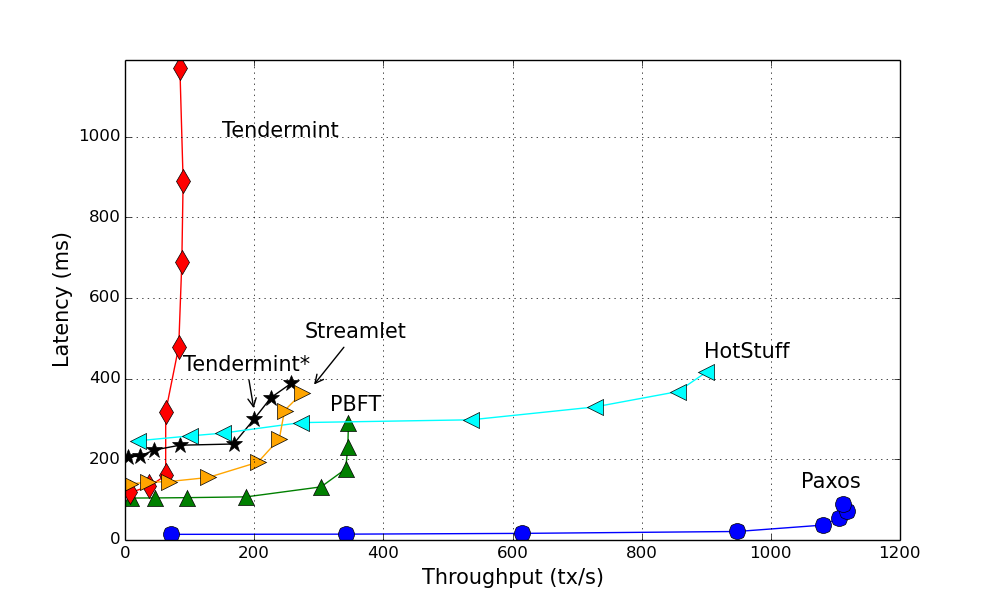}
	\vspace*{-8mm}
	\caption{\textbf{System throughput and the latency on 20-node WAN cluster in Virginia, California, Oregon, and Ohio}}
	\label{fig:WAN_COM.png}
\end{figure}
\vspace*{-5mm}
\subsection{Comparison of Throughput and Latency}
In Figures~\ref{fig:20TC} and~\ref{fig:20TCw}, we discuss the protocols' throughput performance under the same experimental conditions. In both Figures, HotStuff achieves the maximum throughput in LAN and is close to Paxos in WAN deployment. This is due to responsive leader rotation and 4-leader pipelining in HotStuff. In Figure~\ref{fig:20LC}, we explore the average latency performance for all protocols with the same settings. Tendermint latency is the highest due to the $\delta$ wait time. In all protocols, as $\!N\!$ increases, latency increases. This increase is more pronounced for PBFT and Tendermint, because of the all-to-all communication they employ. We also examined the relationship between the system throughput and the latency in WAN and LAN with $\!N\!\!=\!\!20\!$ and 90 clients. The results are shown in Figure~\ref{fig:LvT} and~\ref{fig:WAN_COM.png}. The performance of BFT consensus algorithms is strongly impacted by the number of messages due to tolerance property.

\section{Conclusion and Future Work}

We studied popular deterministic-finality BFT consensus protocols. We analyzed the performance of these protocols, implemented, benchmarked, and evaluated them on AWS under identical conditions. Our results show that the throughput of these protocols do not scale well as the number of participants increases. PBFT and Tendermint suffer the most due to all-to-all communication they employ. HotStuff resolves that problem and shows improved throughput and scalability, comparable to Paxos which only provides crash fault tolerance.

We believe that this work will help developers to choose suitable consensus protocols for their needs. Our findings about the bottlenecks can also pave the way for researchers to design more efficient protocols. As future work, we plan to adopt some bottleneck reduction techniques such as communication relaying nodes~\cite{pigpaxos} and applying them in the considered BFT protocols to improve performance.
\section{ACKNOWLEDGMENTS}
This project is in part sponsored by the National Science Foundation (NSF) under award number CNS-2008243.

\bibliographystyle{IEEEtran}
\bibliography{SurveyBC.bib}
\end{document}